\documentclass[12pt]{article}
\usepackage{graphicx}
\usepackage{epsfig}
\usepackage{graphicx}
\newlength{\vshift}
\newlength{\hshift}

\usepackage{amssymb}
\usepackage{amsmath}
\usepackage{amsthm}
\usepackage{amsopn}
\def\uno{\mbox{1 \kern-.59em {\rm l}}}

\def\beq{\begin{equation}}
\def\eeq{\end{equation}}
\def\bea{\begin{eqnarray}}
\def\eea{\end{eqnarray}}

\begin{document}

 \vspace*{3cm}

\begin{center}

{\bf{\Large  Quantum chaos in quantum dots coupled to bosons}}

\vskip 4em

{ {\bf S. Ahadpour~$^{\dag}$} \footnote{e-mail:ahadpour@uma.ac.ir
}\: and \: {\bf N. Hematpour ~$^{\dag}$
}\footnote{e-mail:N$_{-}$Hemat66@yahoo.com }}

\vskip 1em

$^{\dag}$Department of Physics, University of Mohaghegh Ardabili,
Ardabil, Iran.

 \end{center}

 \vspace*{1.9cm}

\begin{abstract}
Chaos transition, as an important topic, has become an active
research subject in non-linear science. By considering a Dicke
Hamiltonian coupled to a bath of harmonic oscillator, we have been
able to introduce a logistic map with quantum corrections. Some
basic dynamical properties, such as Lyapunov exponents  and
 bifurcation diagram of the model are studied. we show that in this model, the transition from integrable motion to periodic,
  chaotic and hyperchaotic as the control parameter $r$
is increased.
\end{abstract}

PACS number(s): 05.45.+b, 87.23.Cc
%%%%%%%%%%%%%%%%%%%%%%%%%%%%%%%%%%%%%%%%%%%%%%%%%%%%%%%%%%%%%%%%%%%
\newpage
%%%%%%%%%%%%%%%%%%%%%%%%%%%%%%%%%%%%%%%%%%%%%%%%%%%%%
\section{Introduction}
%%%%%%%%%%%%%%%%%%%%%%%%%%%%%%%%%%%%%%%%%%%%%%%%%%%%%
Investigation in the area of quantum dots couple with another quantum dot or one external field  has received a great deal of attention in recent
years. Deterministic equations have been used nowadays in many applications \cite{L. Larger} . For instance, a chaotic system is one which displays
complex and unpredictable behavior.
 These systems with some of properties such as sensitivity to initial condition \cite{M. W. Lee},  system parameter, ergodicity \cite{R. Brown}
and mixing \cite{M. A. Jafarizadeh}. One of the popular fields of quantum
chaos addresses quantum maps. Classical maps exhibit the properties of chaotic dynamical systems.
Various chaotic maps use in Cryptography, Watermarking, Random number generators and lots of
other fields. On the other hand, a big number of researches have been dedicated so far to quantum
maps as paradigms of quantum chaos \cite{M. V. Bersy, A. Lakshminarayan}. M.Goggin et al. in their work, the
quantum correction were made by introducing a periodic noise to a model of oscillatory system
coupled to a bath of oscillators. They derive a new logistic map with quantum computing that illustrate
a period-doubling route to chaos \cite{M. E. Goggin}. The quantization schemes for maps are different and many
classically chaotic maps have been quantized including the standard map \cite{D. L. Shepelyansky, G. Casati}, logistic map
\cite{M. E. Goggin}, baker map on the torus \cite{N. L. Balazs, A. Lakshminarayan1, M. Fannes} and on sphere \cite{P. Pakonski}.
In present paper we introduce  generalized Dick model has been considered to present a new quantum map
which demonstrates positive lyapunov exponents.The comparison between analytically studied and simulation
results indicate that the propose methods provide good estimates.
The paper is organized as follows. In section 2, introduces the model. In section 3, we calculate the lyapunov exponents and KS-entropy.
 Finally, in section 4, we conclude the paper. Respective appendix can be found at the
end of paper.

%*************************************************%%%%%%%%%%%%%%%%%%%%%%%%%%%%%%%%%%%
\section{The model}
%*************************************************%%%%%%%%%%%%%%%%%%%%%%%%%%%%%%%%%%%
We start with the hamiltonian:

\bea
H=a^{\dag}a+\omega_{A}J_{z}+\frac{\gamma}{\sqrt{N}}(a^{\dag}+a)(J_{-}+J_{+})+\frac{\gamma}{\sqrt{N}}V(J_{-},J_{+})\Sigma_{n}\delta(t-nT),
\eea

this hamiltonian is the generalized Dick Hamiltonian. In fact we add a delta function to Dick Hamiltonian. Therefore, $a$ and $a^{\dag}$ bosonic annihilation and creation operators, energy separation of N two-level atoms equal to $\hbar$${\tilde{\omega}}_{A}$ , $\hbar=1$ ,
 field of frequency ${\tilde{\omega}}_{f}$, ${\omega}_{A}={\tilde{\omega}}_{A}/{\tilde{\omega}}_{f} \geq 0$
is given in units of the frequency of the field,
${\gamma}={\tilde{\gamma}}/{\tilde{\omega}}_{f}$ is the coupling
parameter. ${J}_{z}$ the atomic relative population operator, and
${J}_{\pm}$ the atomic transition operators.
 The commutation relations for the fermionic operator J are:

$[{J_{+} ,J_{-}}]=2J_{z}$ , $[J_{z} ,J_{\pm}]=\pm J_{\pm}$ , $[a ,a^{\dag}]=1$.

Heisenberg equation of motion reads:

\bea
\dot{J_{+}}= i[H ,J_{+}],
\eea

 by considering     $  (-i\frac{\gamma}{\sqrt{N}}[J_{+} ,V(J_{+} ,J_{-})]= f(J_{-} ,J_{+}))   $,\\
  we have :
 %\begin{equation}
 $$
\left\{\begin{array}{l}$$\dot{J_{+}}= i\omega_{A}J_{z}- i\frac{\gamma}{\sqrt{N}}(a^{\dag}+a)(2J_{z})+ f(J_{+} ,J_{-})\Sigma_{n}\delta(t-nT),\;\;\;\;\;\;\;(A1)$$\\
\\
\dot{a}= -ia- i\frac{\gamma}{\sqrt{N}}(J_{-}+J_{+}).\;\;\;\;\;\;\;\;\;\;\;\;\;\;\;\;\;\;\;\;\;\;\;\;\;\;\;\;\;\;\;\;\;\;\;\;\;\;\;\;\;\;\;\;\;\;\;\;\;\;\;\;\;\;\;\;(A2)
\end{array}\right.
$$
%\end{equation}

In order to obtain an equation for ${J_{+}}(t)$, we substitute the solution of (A2)  into  (A1):
\bea
\dot{J_{+}}= i\omega_{A}J_{+}- i\frac{\gamma}{\sqrt{N}}a^{\dag}(0)(\exp(it))(2J_{z})- i\frac{\gamma}{\sqrt{N}}a(0)(\exp(-it))(2J_{z})\nonumber\\+ f(J_{+} ,J_{-})\Sigma_{n}\delta(t-nT),
\eea

we integrate of Eq. (3) from t= nT to t= (n+1)T, so as to obtain

\bea
J_{+}((n+1)T)- J_{+}(nT)\exp(i\omega_{A}T)=
\left[-i\frac{\gamma}{\sqrt{N}}a^{\dag}(0){\int_{nT}}^{(n+1)T}(2J_{z})\exp(i(1-\omega_{A})t)dt\right.\nonumber\\
\left. -i\frac{\gamma}{\sqrt{N}}a(0){\int_{nT}}^{(n+1)T}(2J_{z})\exp(-i(1+\omega_{A})t)dt\right]+ f(J_{+}(nT) ,J_{-}(nT))\exp(i\omega_{A}T).
\eea

We assume that $ <a(0)> = 0 $ , $ <a^{\dag}(0)> = 0 $, the expectation value of  $J_{+}$ can be expressed in:

\bea
<J_{+}((n+1)T)> = \left( <J_{+}(nT)> + <f(J_{+}(nT) , J_{-}(nT))>\right)\exp(i\omega_{A}T).
\eea

For comparison our quantum results to the familiar logistic map we choose the ``force":

\bea
f(J_{+(n)} ,J_{-(n)}) = -  J_{+(n)} + r \left(J_{+(n)} - J_{-(n)}J_{+(n)}\right)\exp(-i\omega_{A}T),
\eea

where r is an adjustable parameter. Eq. (5) then becomes :

\bea <J_{+(n+1)}> = r \left( <J_{+(n)}> - < J_{-(n)}J_{+(n)}>
\right), \eea where $x=<J_{+}>=< J_{-}>$ , then: \bea
x_{n+1}=r(x_{n} - {x_{n}}^{2}).\nonumber \eea

Which is the classical logistic map .

The logistic map is one of the most
studied discrete chaotic maps. It was first proposed as pseudo
random number generator by Von Neumann in 1947 partly because it had
a "known algebraic distribution and mentioned later, in 1969, by
Knuth\cite{S. C. Phatak S,C. Peng}. The logistic map is given by:

\begin{equation}
x_{n+1}=r x_{n}(1-x_{n})
\end{equation}
where $x_{n}\in(0,1)$ and $r$ are the system variable and parameter,
respectively, and $n$ is the number of iterations. Thus, given an
initial value $x_{0}$ and a parameter $r$. In this paper, we refer
to $x_{0}$ and $r$ as the initial state of the logistic map.
$$x_{n+1}=r x_{n}(1-x_{n}),\;\;\;\;\; for\;\;\;\;\ x_{n}\in(0,1),\;\;\;and \;\;\;r\in(3.99996,4] $$
and its behavior is full chaotic(Fig 1). The choice of $r$  in the
equation above guarantees the existence of a chaotic orbit that can
be shadowed by only one map as stated in \cite{N. Smaoui}. The
lyapunov exponent h is taken as a measure of the  ``chaoticness" of
N quantum dots coupled with a single bosonic mode. The dynamical
systems, even the discrete-time and one-variable ones have different
types of behaviour. The system can be in a fixed point and nothing
changes, the trajectory of the system may also be a cycle with a
certain period. Fixed point and periodic orbits may be stable or
unstable. We are usually interested in an invariant measure ¹, i.e.
a probability measure that does not change under the dynamics. The
probability measure $\rho(x)=\frac{1}{\pi\sqrt{x(1-x)}}$  on [0; 1]
is a Sinai-Rulle-Bowen (SRB) measure which is an invariant measure
which describes statistically the stationary states of the system
and absolutely continues with respect to Lebesgue measure. Entropy
sometimes is called ''uncertainty" and sometimes ''information". In
order to study the stability, entropy can be used as an acceptable
parameter. Also, by considering SRB measure, it is possible to
continue this study with KS-entropy, the well-known measure for
chaos in dynamical system. This section introduces the KS-entropy
for hierarchy of pair-coupled maps with dynamical parameter. We try
to calculate Lyapunov exponent as another tool to study the
stability.\\
\bea h={\int_{0}}^{1}\ln|F^{'}(x)| \rho(x)dx =
2{\int_{0}}^{1/2}\frac{\ln|r(1-2x)|}{\pi\sqrt{x(1-x)}}dx, \eea with
change of variable $x={\sin}^{2}(\theta)$ ,
$dx=2\sin(\theta)\cos(\theta)d\theta$, and using Eq. (9) becomes
\bea
h=2{\int_{0}}^{\pi/4}\frac{\ln|r(1-2{\sin}^{2}(\theta))|}{\pi\sin(\theta)\cos(\theta)}2\sin(\theta)\cos(\theta)d\theta\nonumber
\eea \bea. \eea\\
 KS-entropy, may also can
be written as:\\
 \bea h=\ln{r}-\ln{2}, \eea for $r=4
\rightarrow h=\ln{2}>0$, it is positive.

For appearance the effect of the quantum correlations by using
$J_{+} = <J_{+}> + \delta J_{+}$ and  $ J_{-} = <J_{-}> + \delta
J_{-}, $ one gets
 \bea
  <J_{+(n+1)}> = r \left(<J_{+(n)}> - <J_{-(n)}> <J_{+(n)}> \right) - r <\delta J_{-} \delta J_{+}>.
 \eea

First, the time derivative of$(\delta J_{+} \delta J_{-})$ is
 \bea {\frac{d}{dt}}(\delta J_{+}
\delta J_{-})=\delta\dot{J_{+}}\delta J_{-}+\delta
J_{+}\delta\dot{J_{-}}. \eea Inserting these two expressions $J_{+}
= <J_{+}> + \delta J_{+}$ and $ J_{-} = <J_{-}> + \delta J_{-} $
into Eq. (6) we end up with\\ \bea f(J_{+} , J_{-})= - <J_{+}> -
\delta J_{+} + \exp(-iw_{A}T)r( <J_{+}> + \delta J_{+}-
<J_{-}><J_{+}>- \nonumber\\\delta J_{-}\delta J_{+}- <J_{-}>\delta
J_{+}- \delta J_{-}<J_{+}> ). \eea

Inserting $\dot{J_{+}}=\delta\dot{J_{+}}$ and
$\dot{J_{-}}=\delta\dot{J_{-}}$ , into Eq. (3) we obtain:

\bea \delta\dot{J_{+}}=iw_{A}(<J_{+}> +\delta J_{+}
)-i\frac{\gamma}{\sqrt{N}}a^{\dag}(0)(\exp(it))[\delta J_{+}\delta
J_{-}-\delta J_{-}\delta J_{+}]-
i\frac{\gamma}{\sqrt{N}}a(0)\nonumber\\(\exp(-it))[ \delta J_{+}
\delta J_{-}-\delta J_{-} \delta J_{+} ]+[ - <J_{+}> - \delta J_{+}
+ \exp(-iw_{A}T)r( <J_{+}> + \delta J_{+}\nonumber\\-
<J_{-}><J_{+}>-\delta J_{-}\delta J_{+}- <J_{-}>\delta J_{+}-\delta
J_{-}<J_{+}> )]\Sigma_{n}\delta(t-nT). \eea

Due to

\bea
\delta\dot{J_{-}}=\delta\dot{J_{+}}^{\dag}=-iw_{A}( <J_{-}> + \delta J_{-} )+i\frac{\gamma}{\sqrt{N}}a(0)(\exp(-it))[\delta J_{-}\delta J_{+}-\delta J_{+}\delta J_{-}]+i\frac{\gamma}{\sqrt{N}}{a^{\dag}}(0)\nonumber\\(\exp(it))[ \delta J_{-} \delta J_{+}-\delta J_{+} \delta J_{-} ]+[ - <J_{-}> - \delta J_{-} + \exp(iw_{A}T)r( <J_{-}> + \delta J_{-}-\nonumber\\<J_{+}><J_{-}>-\delta J_{+}\delta J_{-}- <J_{+}>\delta J_{-}-\delta J_{+}<J_{-}> )]\Sigma_{n}\delta(t-nT),
\eea

 Finally, inserting Eq. (15) and Eq. (16) into Eq. (13), we end up with

\bea
{\frac{d}{dt}}(\delta J_{+} \delta J_{-})=\left[iw_{A}(<J_{+}> +\delta J_{+} )-i\frac{\gamma}{\sqrt{N}}a^{\dag}(0)(\exp(it))[\delta J_{+}\delta J_{-}-\delta J_{-}\delta J_{+}]-i\frac{\gamma}{\sqrt{N}}a(0)\right.\nonumber\\
(\exp(-it))[\delta J_{+}\delta J_{-}-\delta J_{-}\delta J_{+}]+[- <J_{+}> -\delta J_{+}+ \exp(-iw_{A}T)r( <J_{+}> + \delta J_{+}-<J_{-}>\nonumber\\ <J_{+}>-\delta J_{-}\delta J_{+}- <J_{-}>\delta J_{+}-\delta J_{-}<J_{+}>)]\left. \Sigma_{n}\delta(t-nT)\right] {\delta J_{-}}+\delta J_{+}\left[-iw_{A}( <J_{-}> + \delta J_{-} )\right.\nonumber\\
+i\frac{\gamma}{\sqrt{N}}a(0)(\exp(-it))[\delta J_{-}\delta J_{+}-\delta J_{+}\delta J_{-}]+i\frac{\gamma}{\sqrt{N}}{a^{\dag}}(0)(\exp(it))[ \delta J_{-} \delta J_{+}-\delta J_{+} \delta J_{-} ]+[ - <J_{-}>\nonumber\\
 - \delta J_{-} + \exp(iw_{A}T)r( <J_{-}> + \delta J_{-}-<J_{+}><J_{-}>-\delta J_{+}\delta J_{-}-<J_{+}>\delta J_{-}-\delta J_{+}<J_{-}> )]\nonumber\\
\left. \Sigma_{n}\delta (t-nT)\right] . \eea \\
By integrating  Eq. (17), from (nT) to ((n+1)T), and take the
expectation value, by taking into account  $<\delta
J_{-}(nT)>=<\delta J_{+}(nT)>=0$ , $<{a^{\dag}}(0)>=<a(0)>=0$ we
obtain:

\bea <\delta J_{+}((n+1)T)\delta J_{-}((n+1)T)> = -<\delta
J_{+}(nT)\delta J_{-}(nT)>+r\exp(-iw_{A}T)<\delta J_{+}(nT)\delta
J_{-}(nT)>\nonumber \eea \bea -r\exp(-iw_{A}T)<J_{-}><\delta
J_{+}(nT)\delta J_{-}(nT)>-r\exp(-iw_{A}T)<\delta J_{-}(nT)\delta
J_{-}(nT)><J_{+}>+r\nonumber \eea \bea \exp(iw_{A}T)<\delta
J_{+}(nT)\delta J_{-}(nT)>-r\exp(iw_{A}T)<\delta J_{+}(nT)\delta
J_{-}(nT)><J_{+}>-r\exp(iw_{A}T)\nonumber \eea \bea <\delta
J_{+}(nT)\delta J_{+}(nT)><J_{-}>. \eea \\
The calculation of $<\delta J_{+} \delta J_{+}>$  goes as follows:\\
 \bea {\frac{d}{dt}}(\delta
J_{+} \delta J_{+})=\delta\dot{J_{+}}\delta J_{+}+\delta
J_{+}\delta\dot{J_{+}}. \eea appearance the effect of the quantum
correlations by using $J_{+} = <J_{+}> + \delta J_{+}$ and $ J_{-} =
<J_{-}> + \delta J_{-} $ Eq. (6), one gets: \bea f(J_{+} , J_{-})= -
<J_{+}> - \delta J_{+} + \exp(-iw_{A}T)r( <J_{+}> + \delta J_{+}-
<J_{-}><J_{+}>- \nonumber\\\delta J_{-}\delta J_{+}- <J_{-}>\delta
J_{+}- \delta J_{-}<J_{+}> ), \eea

Inserting $\dot{J_{+}}=\delta\dot{J_{+}}$ , into Eq. (3) we obtain:

\bea \delta\dot{J_{+}}=iw_{A}(<J_{+}> +\delta J_{+}
)-i\frac{\gamma}{\sqrt{N}}a^{\dag}(0)(\exp(it))[\delta J_{+}\delta
J_{-}-\delta J_{-}\delta J_{+}]-
i\frac{\gamma}{\sqrt{N}}a(0)\nonumber\\(\exp(-it))[ \delta J_{+}
\delta J_{-}-\delta J_{-} \delta J_{+} ]+[ - <J_{+}> - \delta J_{+}
+ \exp(-iw_{A}T)r( <J_{+}> + \delta J_{+}\nonumber\\-
<J_{-}><J_{+}>-\delta J_{-}\delta J_{+}- <J_{-}>\delta J_{+}-\delta
J_{-}<J_{+}> )]\Sigma_{n}\delta(t-nT), \eea Finally, inserting Eq.
(21) into Eq. (19), we end up with: \bea {\frac{d}{dt}}(\delta J_{+}
\delta J_{+})=[iw_{A}(<J_{+}> +\delta J_{+}
)-i\frac{\gamma}{\sqrt{N}}a^{\dag}(0)(\exp(it))[\delta J_{+}\delta
J_{-}-\delta J_{-}\delta J_{+}]-
i\frac{\gamma}{\sqrt{N}}a(0)\nonumber \eea \bea (\exp(-it))[ \delta
J_{+} \delta J_{-}-\delta J_{-} \delta J_{+} ]+[ - <J_{+}> - \delta
J_{+} + \exp(-iw_{A}T)r( <J_{+}> + \delta J_{+}- <J_{-}>\nonumber
\eea \bea <J_{+}>-\delta J_{-}\delta J_{+}- <J_{-}>\delta
J_{+}-\delta J_{-}<J_{+}> )]\Sigma_{n}\delta(t-nT)]\delta
J_{+}+\delta J_{+}[iw_{A}(<J_{+}> +\nonumber \eea \bea \delta J_{+}
)-i\frac{\gamma}{\sqrt{N}}a^{\dag}(0)(\exp(it))[\delta J_{+}\delta
J_{-}-\delta J_{-}\delta J_{+}]-
i\frac{\gamma}{\sqrt{N}}a(0)\nonumber \eea \bea (\exp(-it))[ \delta
J_{+} \delta J_{-}-\delta J_{-} \delta J_{+} ]+[ - <J_{+}> - \delta
J_{+} + \exp(-iw_{A}T)r( <J_{+}> + \delta J_{+}- <J_{-}>\nonumber
\eea \bea <J_{+}>-\delta J_{-}\delta J_{+}- <J_{-}>\delta
J_{+}-\delta J_{-}<J_{+}> )]\Sigma_{n}\delta(t-nT)]. \eea

By integrating  Eq. (22), from (nT) to ((n+1)T), and take the
expectation value, by taking into account  $<\delta
J_{-}(nT)>=<\delta J_{+}(nT)>=0$ , $<{a^{\dag}}(0)>=<a(0)>=0$ we
obtain:

\bea
<\delta J_{+}((n+1)T) \delta J_{+}((n+1)T)>\exp(-2i\omega_{A}(n+1)T)-<\delta J_{+}(nT) \delta J_{+}(nT)>\exp(-2i\omega_{A}nT)=\nonumber
\eea
\bea
\exp(-2i\omega_{A}nT)(-<\delta J_{+}(nT) \delta J_{+}(nT)>+\exp(-i\omega_{A}T)r(<\delta J_{+}(nT) \delta J_{+}(nT)>-<J_{-}(nT)>\nonumber
\eea
\bea
<\delta J_{+}(nT) \delta J_{+}(nT)>-<J_{+}(nT)><\delta J_{-}(nT) \delta J_{+}(nT)>))+\exp(2i\omega_{A}nT)(-<\delta J_{+}(nT)\nonumber
\eea
\bea
\delta J_{+}(nT)>+\exp(-i\omega_{A}T)r(<\delta J_{+}(nT)\delta J_{+}(nT)>-<J_{-}(nT)><\delta J_{+}(nT) \delta J_{+}(nT)>-\nonumber
\eea
\bea
<\delta J_{+}(nT) \delta J_{-}(nT)><J_{+}(nT)>)).
\eea

By taking into account $<\delta J_{+}(nT) \delta J_{-}(nT)>$=$Y_{n}$
, $<\delta J_{+} \delta J_{+}>$=$Z_{n}$ , $<J_{+}(nT)>$=$X_{n}$ and
inserting these values into Eqs. (12), (18) and (23), we obtain:
$$
\left\{\begin{array}{l}$$

X_{(n+1)} = r ( X_{(n)} - {X_{(n)}}^{2} ) - r Y_{(n)}\;\;\;\;\;\;\;\;\;\;\;\;\;\;\;\;\;\;\;\;\;\;\;\;\;\;\;\;\;\;\;\;\;\;\;\;\;\;\;\;\;\;\;\;\;\;\;\;\;\;\;\;\;\;\;\;\;\;\;\;\;\;\;\;\;(B1)\\
\\
Y_{(n+1)} = -Y_{n}+r\exp(-iw_{A}T)([(1-{X_{(n)}}^{*}\\+
\exp(2iw_{A}T)-X_{(n)}\exp(2iw_{A}T))Y_{n}]\nonumber

-{Z_{n}}^{*}X_{n}-\exp(2iw_{A}T)Z_{n}{X_{n}}{*})\;\;\;\;\;\;\;\;\;\;(B2)\\
\\
Z_{(n+1)} = -Z_{n}\exp(2iw_{A}T)+r\exp(iw_{A}T)(2Z_{n}-2{X_{n}}{*}Z_{n}-X_{n}{Y_{n}}^{*}-Y_{n}X_{n})\;\;(B3).
\end{array}\right.
$$

These equations give us lowest-order quantum corrections.We consider $\beta=iw_{A}T$.

%*************************************************%%%%%%%%%%%%%%%%%%%%%%%%%%%%%%%%%%%
\section{The onset of quantum chaos}
%*************************************************%%%%%%%%%%%%%%%%%%%%%%%%%%%%%%%%%%%
The research on the chaos of Eq. (B) can begin with the analysis on
the stability of the fixed point. Assume $(x, y, z)$ is the fixed
point of
Eq. (B), then $(x, y, z)$ is the solution of the equations below:\\
$$
\left\{\begin{array}{l}$$

 x = r ( x - {x}^{2} ) - r y\;\;\;\;\;\;\;\;\;\;\;\;\;\;\;\;\;\;\;\;\;\;\;\;\;\;\;\;\;\;\;\;\;\;\;\;\;\;\;\;\;\;\;\;\;\;\;\;\;\;\;\;\;\;\;\;\;\;\;\;\;\;\;\;\;(C1)\\
\\
y = -y+r\exp(-\beta)([(1-x\\+
\exp(2\beta)-x\exp(2\beta))y]\nonumber

-zx-\exp(2\beta)zx)\;\;\;\;\;\;\;\;\;\;(C2)\\
\\
z = -z\exp(2\beta)+r\exp(\beta)(2z-2xz-xy-yx)\;\;(C3).
\end{array}\right.
$$

By considering $\exp(\beta)=a$ and solution Eq. (C) we obtain:\\
\newpage
$ x = 0,$\\
$(r - 1)/r,$\\
  $ -(- a^4*r + 2*a^3*r^2 + 2*a^3 - 6*a^2*r + 2*a*r^2 + 2*a - r)/(a^4*r - 4*a^3*r^2 + 6*a^2*r - 4*a*r^2 +
  r),$\\
  \\
 $ y = 0,$\\
$0,$\\

   $(- a^8*r - 2*a^7*r^3 + 6*a^7*r^2 + 2*a^7*r + 2*a^7 + 4*a^6*r^4 - 8*a^6*r^3 - 20*a^6*r - 4*a^6 - 14*a^5*r^3 + 42*a^5*r^2 + 14*a^5*r + 14*a^5 + 8*a^4*r^4 - 16*a^4*r^3 - 54*a^4*r - 8*a^4 - 14*a^3*r^3 + 42*a^3*r^2 + 14*a^3*r + 14*a^3 + 4*a^2*r^4 - 8*a^2*r^3 - 20*a^2*r - 4*a^2 - 2*a*r^3 + 6*a*r^2 + 2*a*r + 2*a - r)/(a^8*r^2 - 8*a^7*r^3 + 16*a^6*r^4 + 12*a^6*r^2 - 56*a^5*r^3 + 32*a^4*r^4 + 38*a^4*r^2 - 56*a^3*r^3 + 16*a^2*r^4 + 12*a^2*r^2 - 8*a*r^3 + r^2),
$\\
\\
 $ z = 0,$\\
   $ 0,$\\
   $(2*a^9*r^2 + 4*a^8*r^4 - 8*a^8*r^3 - 4*a^8*r^2 - 8*a^8*r - 16*a^7*r^3 + 48*a^7*r^2 + 16*a^7*r + 8*a^7 + 12*a^6*r^4 - 24*a^6*r^3 + 4*a^6*r^2 - 88*a^6*r - 16*a^6 - 32*a^5*r^3 + 92*a^5*r^2 + 32*a^5*r + 48*a^5 + 12*a^4*r^4 - 24*a^4*r^3 + 4*a^4*r^2 - 88*a^4*r - 16*a^4 - 16*a^3*r^3 + 48*a^3*r^2 + 16*a^3*r + 8*a^3 + 4*a^2*r^4 - 8*a^2*r^3 - 4*a^2*r^2 - 8*a^2*r + 2*a*r^2)/(a^10*r^2 - 8*a^9*r^3 + 16*a^8*r^4 + 13*a^8*r^2 - 64*a^7*r^3 + 48*a^6*r^4 + 50*a^6*r^2 - 112*a^5*r^3 + 48*a^4*r^4 + 50*a^4*r^2 - 64*a^3*r^3 + 16*a^2*r^4 + 13*a^2*r^2 - 8*a*r^3 + r^2).
$
\\
\\

In the thermodynamic limit $\beta \rightarrow 0 $ or $a=1$, we have following fixed points:\\
\\
$(x_{0},y_{0},z_{0})=(0,0,0)$
\\
$(x_{1},y_{1},z_{1})=((r - 1)/r,0,0)$
\\
$(x_{2},y_{2},z_{2})=(-(r^2 - 3*r + 2)/(4*r - 3*r^2),(2*r^4 - 10*r^3
+ 18*r^2 - 14*r + 4)/(9*r^4 - 24*r^3 + 16*r^2),(r^4 - 6*r^3 + 13*r^2
- 12*r + 4)/(9*r^4 - 24*r^3 + 16*r^2))$\\
\\
\\
We substitute $X_{(n+1)}=X$, $X_{n}=x$, $Y_{(n+1)}=Y$, $Y_{n}=y$,
$Z_{(n+1)}=Z$ and $Z_{(n+1)}=z$ into
Eq. (B) obtain:\\

$$
\left\{\begin{array}{l}$$

X = r ( x - {x}^{2} ) - r y\;\;\;\;\;\;\;\;\;\;\;\;\;\;\;\;\;\;\;\;\;\;\;\;\;\;\;\;\;\;\;\;\;\;\;\;\;\;\;\;\;\;\;\;\;\;\;\;\;\;\;\;\;\;\;\;\;\;\;\;\;\;\;\;\;(D1)\\
\\
Y = -y+r\exp(-\beta)([(1-x\\+
\exp(2\beta)-x\exp(2\beta))y]\nonumber

-zx-\exp(2\beta)zx)\;\;\;\;\;\;\;\;\;\;(D2)\\
\\
Z = -z\exp(2\beta)+r\exp(\beta)(2z-2xz-xy-yx)\;\;(D3).
\end{array}\right.
$$
The corresponding bifurcation diagram of state x with respect to $r$
is giving in Fig. 2.

In order to calculate the lyapunov exponents at $x=x_{0}$, $y=y_{0}$
and $z=z_{0}$, we need to calculate the characteristic
roots of the matrix:\\

\begin{equation}
 A=\left[
\begin{array}{ccc}  \frac{\partial X}{\partial
x} & \frac{\partial X}{\partial y} &
\frac{\partial X}{\partial z} \\
\frac{\partial Y}{\partial x} & \frac{\partial Y}{\partial y} &
\frac{\partial
Y}{\partial z} \\
\frac{\partial Z}{\partial x} & \frac{\partial Z}{\partial y} &
\frac{\partial Z}{\partial z}
\end{array} \right ]=\\
\left[
\begin{array}{ccc}  r(1-2x_{0}) & -r  & 0 \\ \frac{r}{a}(y_{0}+z_{0})(-1-a^{2}) & -1+\frac{r}{a}(1-x_{0})(1+a^{2})  & \frac{r}{a}x_{0}(-1-a^{2}) \\
2ar(-z_{0}-y_{0}) & -2rax_{0} & -a^{2}+2ra(1-x_{0})
\end{array} \right ]
\end{equation}

Substituting $x_{0}=0$, $y_{0}=0$ and $z_{0}=0$ , into A:\\

\begin{equation}
  A =\left[
\begin{array}{cccc}  r & -r  & 0 \\ 0 & -1+\frac{r}{a}(1+a^{2})  & 0 \\
0 & 0 & -a^{2}+2ra
\end{array} \right ],
\end{equation}
 and obtain three eigenvalues as below:\\

\begin{equation}
\left|\begin{array}{ccc} r-\lambda & -r & 0 \\
0 & -1+\frac{r}{a}(1+a^{2})-\lambda & 0 \\ 0 & 0 & -a^{2}+2ar-\lambda
\end{array}\right|=(r-\lambda)(-1+\frac{r}{a}(1+a^{2})-\lambda)(-a^{2}+2ar-\lambda)=0
\end{equation}

\bea
{\lambda}_{1}=r , {\lambda}_{2}=-1+\frac{r}{a}(1+a^{2}) , {\lambda}_{3}=-a^{2}+2ar
\eea

for $a=1$, $  {\lambda}_{1}=r , {\lambda}_{2}=-1+2r , {\lambda}_{3}=-1+2r $.\\
\\
As increased $r$, the stability of system (D) is summarized as
follows:\\
(1)\;\;$r=0$, \;$\lambda_{1}=0$, \;$\lambda_{2}=\lambda_{3}< 0$, \;system $(D)$ is periodic.\\
(2)\;\;$0 < r < 0.5$, \;$\lambda_{1}> 0$, \;$\lambda_{2}=\lambda_{3}< 0$, \;the stability of system $(D)$ is chaotic.\\
(3)\;\;$r > 0.5$, \;$\lambda_{1}> 0$, \;$\lambda_{2}=\lambda_{3}> 0$, \;the stability of system $(D)$ is hyperchaotic \cite{O.E. Rossler}.\\
\\
Obviously, the coupled system is ergodic at $(x=x_{0}, y=y_{0},
z=z_{0})$  according to Pessin theorem \cite{JP. Eckman,YaB.
Pessin}. Due to ergodcity of one-dimensional map,
$x_{n+1}=rx_{n}(1-x_{n})$, we have $\lambda_{1}$,  $\lambda_{2}$,
$\lambda_{3}$, and the KS-entropy is equal to sum of positive
Lyapunov exponents.

Now comparing the KS-entropy with sum of Lyapunov exponents, one can show that:\\

\bea h_{KS}={\lambda}_{1}+{\lambda}_{2}+{\lambda}_{3}=5r-2. \eea
\\
For $r=0.4$ \\
 $h_{KS}=0$,\\
therefore the system is not chaotic. The measurable dynamical system
is chaotic for $h_{KS} > 0$  and predictive for  $h_{KS}=0$. For $r
> 0.4$, positive KS-entropy  occurs. For each $r > 1$, set
${\lambda}_{2} = {\lambda}_{3} > {\lambda}_{1}$.
By considering $0.5 < r < 1$ $\Rightarrow$ ${\lambda}_{1} > {\lambda}_{2} = {\lambda}_{3} > 0$.\\

The corresponding bifurcation diagram of state x with respect to $r$
is giving in Fig. 3.
Substituting $x=\frac{r-1}{r}$, $y=0$ and $z=0$ we into A:\\

\begin{equation}
 A =\left[
\begin{array}{cccc}  -r+2 & -r  & 0 \\ 0 & -1+\frac{1}{a}+a  & \frac{r-1}{a} \\
0 & -2a(r-1) & -a^{2}+2a
\end{array} \right ]
\end{equation}

 and obtain three eigenvalues as below:\\

\begin{equation}
{\left|\begin{array}{ccc} -r+2-\lambda & -r & 0 \\
0 & -1+\frac{1}{a}+a-\lambda & \frac{r-1}{a} \\ 0 & -2a(r-1) & -a^{2}+2a-\lambda
\end{array}\right|}=\\\nonumber(-r+2-\lambda)(-1+\frac{1}{a}+a-\lambda)(-a^{2}+2a-\lambda)=0
\end{equation}
\bea
{\lambda}_{1}=-r+2 , {\lambda}_{2}=-1+\frac{1}{a}+a , {\lambda}_{3}=-a^{2}+2a
\eea

for $a=1$, $  {\lambda}_{1}=2-r , {\lambda}_{2}=1 , {\lambda}_{3}=1 $.\\
\\
As increased $r$, the stability of system (D) is summarized as
follows:\\
(1)\;\;$r < 2$, \;$\lambda_{1}> 0$, \;$\lambda_{2}=\lambda_{3}=1> 0$, \;system $(D)$ is hyperchaotic.\\
(2)\;\;$r = 2$, \;$\lambda_{1}= 0$, \;$\lambda_{2}=\lambda_{3}=1> 0$, \;the stability of system $(D)$ is hyperchaotic.\\
(3)\;\;$r > 2$, \;$\lambda_{1}< 0$, \;$\lambda_{2}=\lambda_{3}=1> 0$, \;the stability of system $(D)$ is hyperchaotic.\\
\\

Now comparing the KS-entropy with sum of Lyapunov exponents, one can show that:\\
\bea h_{KS}={\lambda}_{1}+{\lambda}_{2}+{\lambda}_{3}=4-r. \eea
\\
For $r=4$ \\
 $h_{KS}=0$,\\
therefore the system is not chaotic. The measurable dynamical system
is chaotic for $h_{KS} > 0$  and predictive for  $h_{KS}=0$. For $r
< 4$, positive KS-entropy  occurs. For each $r > 1$, set
${\lambda}_{2} = {\lambda}_{3} > {\lambda}_{1}$.
By considering $ r < 1$ $\Rightarrow$ ${\lambda}_{1} > {\lambda}_{2} = {\lambda}_{3} > 0$.\\
The corresponding bifurcation diagram of state x with respect to
$\beta$ is giving in Fig. 5. In accordance with this figure, the
system is chaotic for positive $\beta$.

%%%%%%%%%%%%%%%%%%%%%%%%%%%%%%%%%%%%%%%%%%%%%%%%
%\section{Results}
%%%%%%%%%%%%%%%%%%%%%%%%%%%%%%%%%%%%%%%%%%%%%%%
%In this article, a new chaotic map were offered. It can answer the need of many scientists and researchers
%precise analysis of chaos in a system of array of N quantum dots with a single bosonic mode which is crucial in
%many branches of science such as cryptography.
%Advantage this method our new chaotic map because of
%its nature it has the highest ability to apply in cryptography.
%%%%%%%%%%%%%%%%%%%%%%%%%%%%%%%%%%%%%%%%%%%%%%%%
\section{Conclusion}
%%%%%%%%%%%%%%%%%%%%%%%%%%%%%%%%%%%%%%%%%%%%%%%
In this article, a new scheme for studying the effects of quantum
correlations on a Dicke Hamiltonian coupled to a bath of harmonic
oscillator has been presented. By calculating Some basic dynamical
properties, such as Lyapunov exponents  and bifurcation diagram of
the model, the proposed scheme is proved to be the transition from
integrable motion to periodic, chaotic and hyperchaotic as the
control parameter $r$ is increased.

Further, we do hope that our obtained results through this paper
will pave the way for further studies on quantum chaos.
%%%%%%%%%%%%%%%%%%%%%%%%%%%%%%%%%%%%%%%%%%%%%%%%
\section{Appendix}
%%%%%%%%%%%%%%%%%%%%%%%%%%%%%%%%%%%%%%%%%%%%%%%
Here we derive the Lyapunov exponents of chaotic maps by using fixed point $(x_{2},y_{2},z_{2})$.
For $a=1$, we have following fixed points:\\
\\
$(x_{2},y_{2},z_{2})=(-(r^2 - 3*r + 2)/(4*r - 3*r^2),(2*r^4 - 10*r^3
+ 18*r^2 - 14*r + 4)/(9*r^4 - 24*r^3 + 16*r^2),(r^4 - 6*r^3 + 13*r^2
- 12*r + 4)/(9*r^4 - 24*r^3 + 16*r^2))$\\
\\
for $r=4$,\\
 $(x_{2},y_{2},z_{2})=(0.19,0.105,0.352)$.\\
 \\
The corresponding bifurcation diagram of state $x$ with respect to $r$
is giving in Fig. 4.
Substituting $x=0.19$, $y=0.105$ and $z=0.352$ we into A:\\

\begin{equation}
 A =\left[
\begin{array}{cccc}  2.48 & -4  & 0 \\ -3.656 & 5.48  & -1.52 \\
-3.656 & -1.52 & 5.48
\end{array} \right ]
\end{equation}
and obtain three eigenvalues as below:\\

\begin{equation}
{\left|\begin{array}{ccc} 2.48-\lambda & -4 & 0 \\
-3.656 & 5.48-\lambda & -1.52 \\ -3.656 & -1.52 & 5.48-\lambda
\end{array}\right|}=\\\nonumber(-\lambda^{3}+13.44\lambda^{2}-40.276\lambda-33.618)=0
\end{equation}
\\
${\lambda}_{1}$=$  24919/(3750*(- 2138981/125000 + (10298398109^(1/2)*16875000000^(1/2)*i)/16875000000)^(1/3)) + (- 2138981/125000 + (10298398109^(1/2)*16875000000^(1/2)*i)/16875000000)^(1/3) + 112/25$\\\nonumber
\\\nonumber
${\lambda}_{2}$=$  112/25 - (- 2138981/125000 + (10298398109^(1/2)*16875000000^(1/2)*i)/16875000000)^(1/3)/2 - 24919/(7500*(- 2138981/125000 + (10298398109^(1/2)*16875000000^(1/2)*i)/16875000000)^(1/3)) + (3^(1/2)*(24919/(3750*(- 2138981/125000 + (10298398109^(1/2)*16875000000^(1/2)*i)/16875000000)^(1/3)) - (- 2138981/125000 + (10298398109^(1/2)*16875000000^(1/2)*i)/16875000000)^(1/3))*i)/2$\\\nonumber
\\\nonumber
${\lambda}_{3}$=$  112/25 - (- 2138981/125000 + (10298398109^(1/2)*16875000000^(1/2)*i)/16875000000)^(1/3)/2 - 24919/(7500*(- 2138981/125000 + (10298398109^(1/2)*16875000000^(1/2)*i)/16875000000)^(1/3)) - (3^(1/2)*(24919/(3750*(- 2138981/125000 + (10298398109^(1/2)*16875000000^(1/2)*i)/16875000000)^(1/3)) - (- 2138981/125000 + (10298398109^(1/2)*16875000000^(1/2)*i)/16875000000)^(1/3))*i)/2$\\\nonumber

%%%%%%%%%%%%%%%%%%%%%%%%%%%%%%%%%%%%%%%%%%%%%%%%%%%%%%%%%%%%%%%%%%%%%%%%%%%%%%%%

\newpage
\begin{figure}
\begin{center}
\scalebox{0.50}{\includegraphics{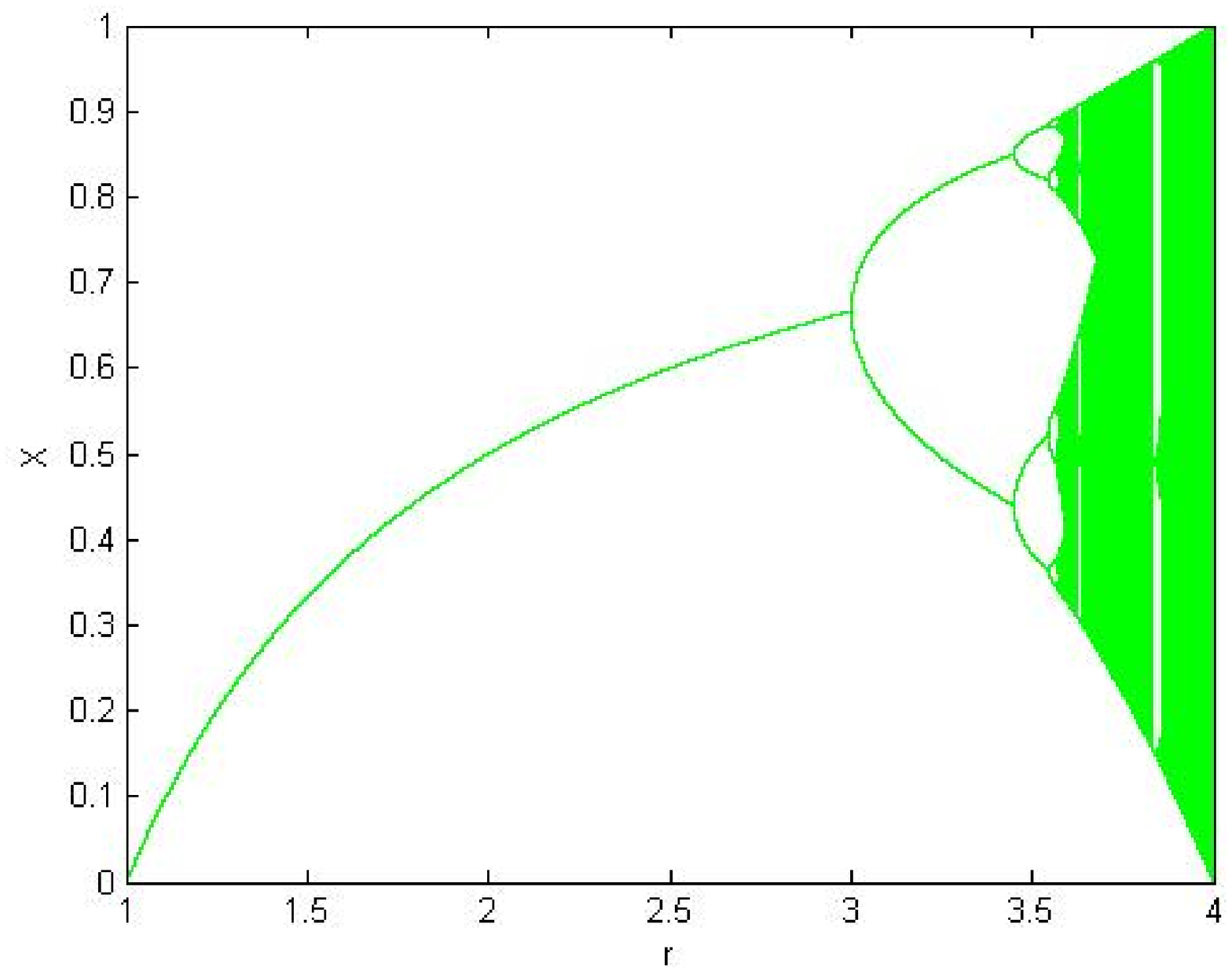}} \caption{{ Bifurcation
diagram for the logistic map, $x_{n+1}=rx_{n}(1-x_{n})$ when the
parameter $r$ is varied from 1 to 4. }\label{fig:1}}
\end{center}
\end{figure}

\begin{figure}
\begin{center}
\scalebox{0.50}{\includegraphics{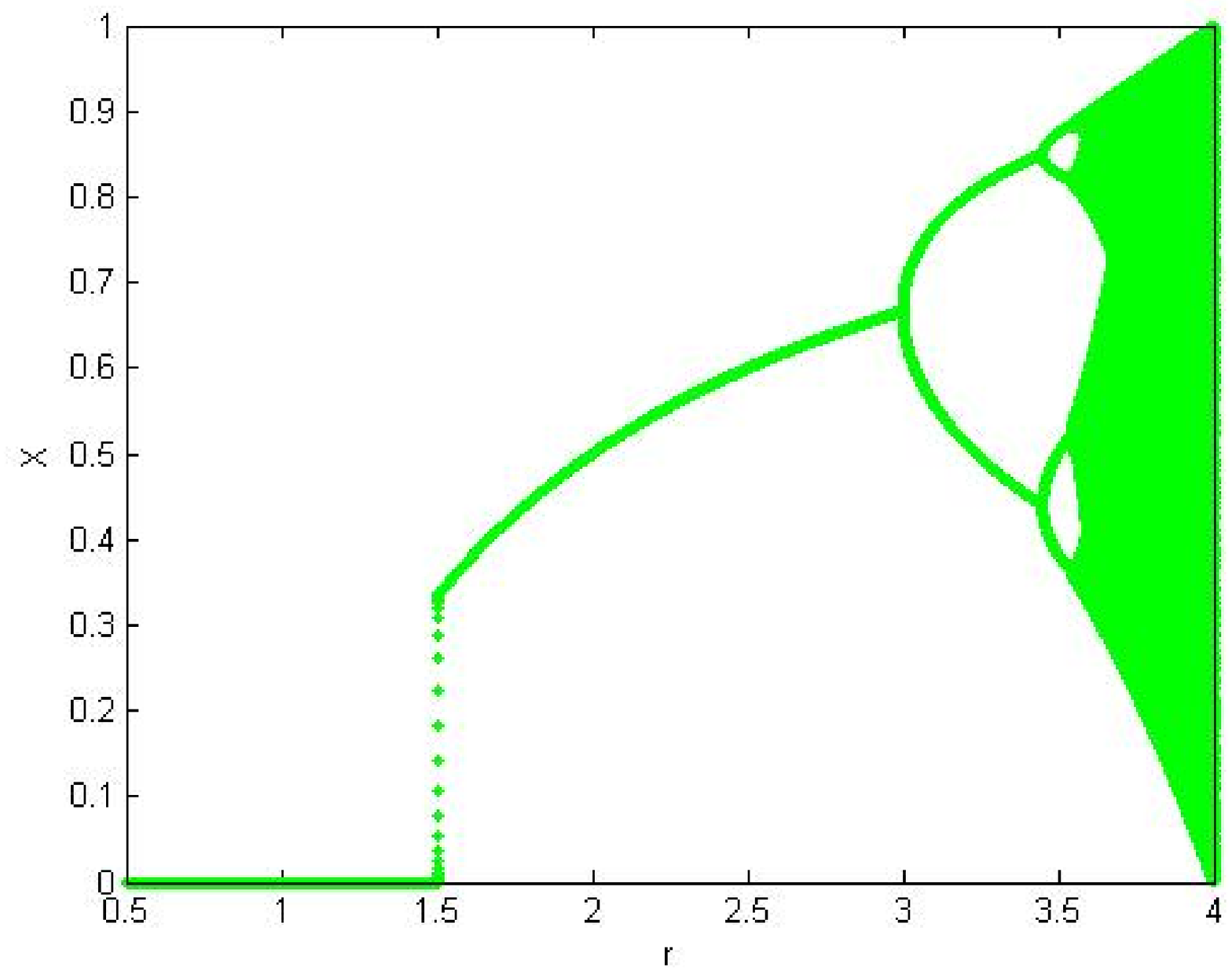}} \caption{{Bifurcation
diagram for increasing $r$, with $a=1$, $x_{0}=0$, $y_{0}=0$ and
$z_{0}=0$.}\label{fig:1}}
\end{center}
\end{figure}

\begin{figure}
\begin{center}
\scalebox{0.50}{\includegraphics{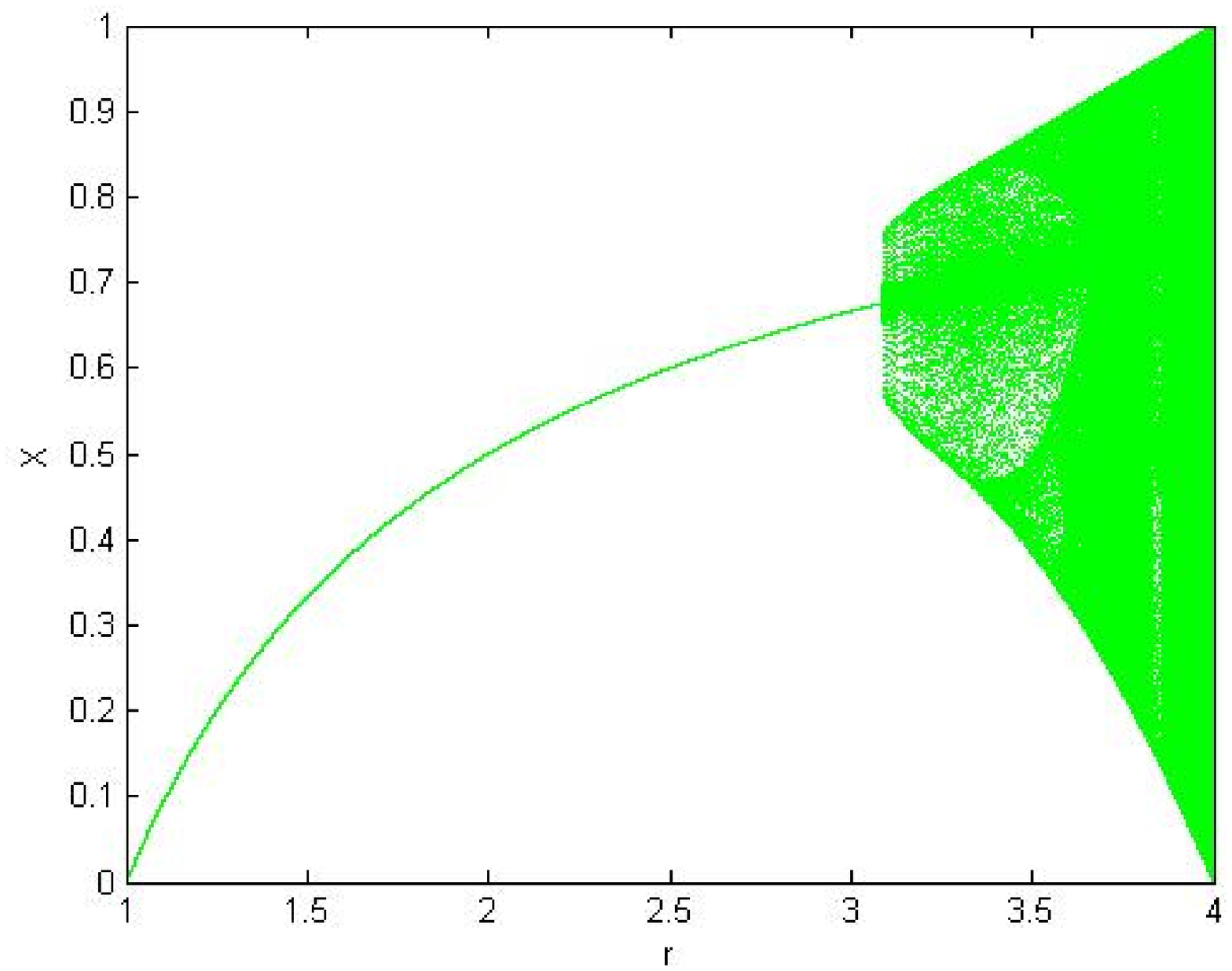}} \caption{{Bifurcation
diagram for increasing $r$, with $a=1$, $x_{0}=\frac{r-1}{r}$, $y_{0}=0$ and
$z_{0}=0$.}\label{fig:1}}
\end{center}
\end{figure}

\begin{figure}
\begin{center}
\scalebox{0.50}{\includegraphics{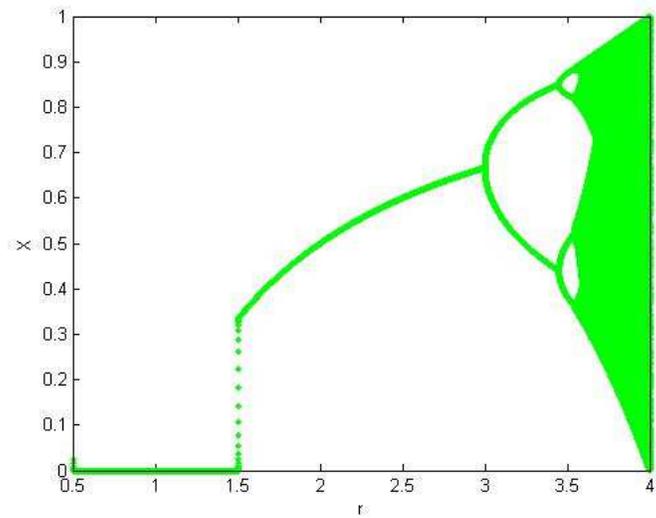}} \caption{{Bifurcation
diagram for $r=4$, with $a=1$, $x_{0}=0.19$, $y_{0}=0.105$ and
$z_{0}=0.352$.}\label{fig:1}}
\end{center}
\end{figure}

\begin{figure}
\begin{center}
\scalebox{0.50}{\includegraphics{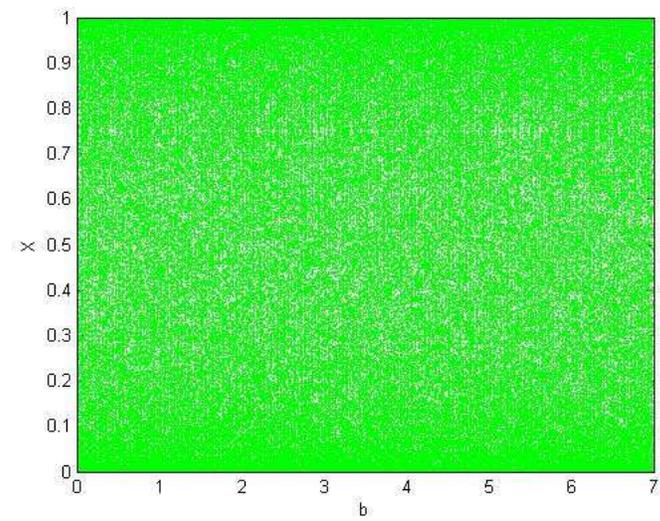}} \caption{{Bifurcation
diagram for increasing $\beta$, with $r=4$, $x_{0}=0.2$, $y_{0}=0$ and
$z_{0}=0$.}\label{fig:1}}
\end{center}
\end{figure}

\end{document}